\documentstyle[12pt,aps]{revtex}

\widetext
\draft
\tighten
\begin{document}

\title{{\normalsize {\rm {\bf FUNDAMENTOS MATEM\'{A}TICOS DE TEOR\'{I}AS SOBRE
ESPACIOS NOCONMUTATIVOS}}\thanks{
Ser\'{a} publicado en las Memorias del VII Reuni\'{o}n Nacional
Acad\'{e}mica de F\'{i}sica y Matem\'{a}ticas, ESFM IPN, D. F., M\'{e}xico,
13-17 de Mayo, 2002}}}

\author{{\bf Valeri V. Dvoeglazov}}

\address{Universidad de Zacatecas\\
Apartado Postal 636, Suc. UAZ\\
Zacatecas 98062, Zac., M\'exico\\
E-mail: valeri@ahobon.reduaz.mx\\
URL: http://ahobon.reduaz.mx/\~\,valeri/valeri.htm}

\date{May 13, 2002}
\maketitle

\begin{abstract}
\hspace*{-10mm} {\bf RESUMEN1.} (Para ArXivos de LANL) En la pl\'atica
investigamos la cuesti\'on de conmutaci\'on de derivadas parciales cuando
la funci\'on sujeta a diferenciaci\'on tiene dependencia
tanto expl\'{\i}cita como impl\'{\i}cita. Aplicamos los resultados
a teor\'{\i}as cu\'anticas.

\hspace*{-7mm} {\bf RESUMEN2.} Recientemente se han encontrado problemas en
la definici\'{o}n de la derivada parcial en el caso de presencia de
dependencia, tanto expl\'{\i}cita como impl\'{\i}cita, de la funci\'{o}n
sujeta a diferenciaci\'{o}n en el an\'{a}lisis cl\'{a}sico. En la presente
pl\'{a}tica investigamos la influencia de este descubrimiento a la mec\'{a}nica
cu\'{a}ntica y teor\'{\i}as de campos cl\'asicos y cu\'{a}nticos.
Sorprendentemente, ciertos conmutadores de los operadores de las coordenadas
de espacio-tiempo {\it no} son iguales a cero. Entonces, proveemos la base
para modernas teor\'{\i}as noconmutativas.
\end{abstract}

\pagestyle{empty}

\thispagestyle{empty}

\bigskip

\thispagestyle{empty}

\bigskip

\noindent Es muy conocido que la idea que los operadores de 4-coordenadas
{\it no} conmutan $[\hat{x}_{\mu },\hat{x}_{\nu }]_{-}\neq 0$ ha sido
propuesta por H. Snyder~\cite{snyder}. En algunas teor\'{\i}as basadas en
este postulado, la simetr\'{i}a de Lorentz aparece ser rota. Recientemente,
mucho inter\'{e}s ha sido atra\'{\i}do a esta idea~\cite{noncom,kruglov} en
el \'{a}mbito de teor\'{i}as ``brane''.

Adem\'{a}s, el famoso procedimiento de la derivaci\'{o}n de dos ecuaciones
de Maxwell ($div{\bf H}=0$ y $\partial {\bf H}/\partial t+curl{\bf E}=0$)
por Feynman y Dyson~\cite{FD} contiene intr\'{\i}nsecamente la
no-conmutatividad de velocidades $[\dot{x}_{i}(t),\dot{x}_{j}(t)]_{-}={\frac{
i\hbar }{m^{2}}}\epsilon _{ijk}B_{k}$ que tambi\'{e}n puede ser considerada
como una contradicci\'{o}n con las teor\'{\i}as bien aceptadas. Dyson escribe
muy inteligentemente: ``Feynman en 1948 no fue la \'{u}nica persona que
trat\'{o} de construir teor\'{\i}as fuera de la rama de f\'{\i}sica convencial...
Todos estos programas radicales, incluyendo aquel de Feynman,
fracasaron...Yo estoy en desacuerdo con Feynman ahora, como frecuentamente
fui cuando el estaba vivo...''

De otro modo, se ha descubierto que el concepto de derivada parcial {\it no}
es bien definido en el caso de la dependencia, tanto expl\'{\i}cita como
impl\'{\i}cita, de la funci\'{o}n correspondiente, en la que las derivadas
act\'{u}an~\cite{chub,brown} (v\'{e}ase tambi\'{e}n la discusi\'{o}n en~\cite
{chja}). El muy conocido ejemplo de esta situaci\'{o}n es el problema del
campo de carga acelerada~\cite{landau}.\footnote{
Primero de todo, Landau y Lifshitz escribieron que las funcciones dependan
de $t^{\prime }$ y solo por $t^{\prime }+R(t^{\prime })/c=t$ las dependan
impl\'{\i}citamente de $x,y,z,t$. Sin embargo, mas adelante (cuando calcularon la
formula (63.7)) utilizaron la dependencia explicita de $R$ de los
coordenadas espaciales del punto de observaci\'{o}n. Chubykalo y Vlayev
declararon que la derivada en tiempo y $curl$ {\it no} conmutan en su caso
de consideraci\'{o}n del problema. De hecho, Jackson fue desacuerdo con esta
declaraci\'{o}n, basando su opini\'{o}n en las definiciones (``las
ecuaciones que representan ley de Faraday y la ausencia de cargas magneticas
...se satisfacen automaticamente''; v\'ease su Introducci\'{o}n en~[6c]). Sin
embargo, el esta de acuerdo con~\cite{landau} que uno tiene que encontrar
``una contribuci\'{o}n al derivada parcial para tiempo fijado $t$ de
dependencia explicita de las coordenadas espaciales (del punto de
observaci\'{o}n).'' \v{S}kovrlj y Ivezi\'{c}~[6d] nombraron esta derivada parcial
como `derivada parcial {\it completa}'; Chubykalo y Vlayev~[6a], como
`derivada {\it total} con respeto de dada variable'; el termino sugerido por
Brownstein~\cite{brown} es `la derivada {\it entera-parcial}'. De hecho, Chubykalo
y Vlayev declararon que sin tomar en cuenta la dependencia expl\'{\i}cita de los
campos electromagn\'eticos las ecuaciones de Maxwell {\it no} satisfacen. No
esta bien claro para mi que intentaba dudar Prof. Jackson en su
trabajo~[6c]. Las funciones de campo y potenciales aparecen ser las funciones
del tipo ${\bf E}({\bf x},t^{\prime }({\bf x},t))$}

Las propiedades de una derivada entera-parcial son sorprendentes. Nos
permitimos estudiar el caso cuando hay dependencias expl\'{\i}cita e
impl\'{\i}cita $f({\bf p},E({\bf p}))$. Es muy conocido que la energ\'{\i}a en
relativismo est\'{a} conectada con el 3-momento como $E=\pm \sqrt{{\bf p}
^{2}+m^{2}}$; se utiliza el sistema de unidades $c=\hbar =1$. En otras
palabras, debemos escoger el hiperboloido 3-dimensional de todo espacio de
Minkowski, y la energ\'{\i}a ya {\it no} es una cantidad independiente.
Entonces, las soluciones de las ecuaciones relativistas son, de hecho, las
funciones que contienen ambos tipos de dependencia, expl\'{\i}cita e
impl\'{\i}cita. Nos permitimos calcular el conmutador de la derivada entera-parcial
$\hat{\partial}/\hat{\partial}p_{i}$
y $\hat{\partial}/\hat{\partial}E$.\footnote{
Para hacer una distincci\'{o}n entre derivando la funcci\'{o}n expl\'{\i}cita y
aquella que contiene como dependencia expl\'{\i}cita tanto impl\'{\i}cita, la derivada
entera-parcial puede ser denotada como $\hat{\partial}$.} En el caso
general, uno tiene
\begin{equation}
{\frac{\hat{\partial}f({\bf p},E({\bf p}))}{\hat{\partial}p_{i}}}\equiv {
\frac{\partial f({\bf p},E({\bf p}))}{\partial p_{i}}}+{\frac{\partial f(
{\bf p},E({\bf p}))}{\partial E}}{\frac{\partial E}{\partial p_{i}}}\,.
\end{equation}
Aplicando esta regla, encontramos sorprendentemente que
\begin{eqnarray}
&&[{\frac{\hat{\partial}}{\hat{\partial}p_{i}}},{\frac{\hat{\partial}}{\hat{
\partial}E}}]_{-}f({\bf p},E({\bf p}))={\frac{\hat{\partial}}{\hat{\partial}
p_{i}}}{\frac{\partial f}{\partial E}}-{\frac{\partial }{\partial E}}({\frac{
\partial f}{\partial p_{i}}}+{\frac{\partial f}{\partial E}}{\frac{\partial E
}{\partial p_{i}}})=  \nonumber \\
&=&{\frac{\partial ^{2}f}{\partial E\partial p_{i}}}+{\frac{\partial ^{2}f}{
\partial E^{2}}}{\frac{\partial E}{\partial p_{i}}}-{\frac{\partial ^{2}f}{
\partial p_{i}\partial E}}-{\frac{\partial ^{2}f}{\partial E^{2}}}{\frac{
\partial E}{\partial p_{i}}}-{\frac{\partial f}{\partial E}}{\frac{\partial
}{\partial E}}({\frac{\partial E}{\partial p_{i}}})\,.  \label{com}
\end{eqnarray}
Entonces, si $E=\pm \sqrt{m^{2}+{\bf p}^{2}}$ y $\partial E/\partial
p_{i}=p_{i}/E$ uno tiene que la expresi\'{o}n (\ref{com}) ha de ser igual a
$(p_{i}/E^{2}){\frac{\partial f({\bf p},E({\bf p}))}{
\partial E}}$.\footnote{
Recuerda que suponemos que la dependencia de $\partial E/\partial p_{i}$
es la misma que la de  funci\'{o}n $f$ sujeta de derivaci\'{o}n.} Con la
libertad de elecci\'{o}n de la normalizaci\'{o}n, el coeficiente es el
campo el\'{e}ctrico longitudinal en el base de helicidad (los campos
el\'{e}ctricos y magn\'{e}ticos pueden ser derivados de los
4-potenciales que se presentaron en~\cite{hb}).\footnote{Recuerda que en
el base escogida tenemos:
\begin{mathletters} \begin{eqnarray} &&\epsilon
_{\mu }({\bf p},\lambda =+1)={\frac{1}{\sqrt{2}}}{\frac{e^{i\phi }}{
p}}\pmatrix{ 0, {p_x p_z -ip_y p\over \sqrt{p_x^2 +p_y^2}}, {p_y p_z +ip_x
p\over \sqrt{p_x^2 +p_y^2}}, -\sqrt{p_x^2 +p_y^2}}\,, \\
&&\epsilon _{\mu }({\bf p},\lambda =-1)={\frac{1}{\sqrt{2}}}{\frac{e^{-i\phi }
}{p}}\pmatrix{ 0, {-p_x p_z -ip_y p\over \sqrt{p_x^2 +p_y^2}}, {-p_y p_z
+ip_x p\over \sqrt{p_x^2 +p_y^2}}, +\sqrt{p_x^2 +p_y^2}}\,, \\
&&\epsilon _{\mu }({\bf p},\lambda =0)={\frac{1}{m}}\pmatrix{ p, -{E \over p}
p_x, -{E \over p} p_y, -{E \over p} p_z }\,, \\
&&\epsilon _{\mu }({\bf p},\lambda =0_{t})={\frac{1}{m}}\pmatrix{E , -p_x,
-p_y, -p_z }\,.
\end{eqnarray}
\end{mathletters}
Y,
\begin{mathletters}
\begin{eqnarray}
&&{\bf E}({\bf p},\lambda =+1)=-{\frac{iEp_{z}}{\sqrt{2}pp_{l}}}{\bf p}-{\frac{
E}{\sqrt{2}p_{l}}}\tilde{{\bf p}},\quad {\bf B}({\bf p},\lambda =+1)=-{\frac{
p_{z}}{\sqrt{2}p_{l}}}{\bf p}+{\frac{ip}{\sqrt{2}p_{l}}}\tilde{{\bf p}}, \\
&&{\bf E}({\bf p},\lambda =-1)=+{\frac{iEp_{z}}{\sqrt{2}pp_{r}}}{\bf p}-{\frac{
E}{\sqrt{2}p_{r}}}\tilde{{\bf p}}^{\ast },\quad {\bf B}({\bf p},\lambda
=-1)=-{\frac{p_{z}}{\sqrt{2}p_{r}}}{\bf p}-{\frac{ip}{\sqrt{2}p_{r}}}\tilde{
{\bf p}}^{\ast }, \\
&&{\bf E}({\bf p},\lambda =0)={\frac{im}{p}}{\bf p},\quad {\bf B}({\bf p}
,\lambda =0)=0,
\end{eqnarray}
\end{mathletters}
with $\tilde{{\bf p}}=\pmatrix{p_y\cr -p_x\cr -ip\cr}$.}
De otro modo, el conmutador
\begin{equation}
\lbrack {\frac{\hat{\partial}}{\hat{\partial}p_{i}}},{\frac{\hat{\partial}}{
\hat{\partial}p_{j}}}]_{-}f({\bf p},E({\bf p}))={\frac{1}{E^{3}}}{\frac{
\partial f({\bf p},E({\bf p}))}{\partial E}}[p_{i},p_{j}]_{-}\,.
\end{equation}
Esto puede ser considerado que sea igual a zero si nosotros no vamos a creer
en el genio de Feynman. El ha postulado que el conmutador de las velocidades
(o, claro, de los 3-momentos) es igual a $[p_{i},p_{j}]\sim i\hbar \epsilon
_{ijk}B^{k}$, i.e., al campo magn\'{e}tico.\footnote{De hecho, si vamos poner
en  corespondencia a los  momentos sus operadores de mec\'anica
cu\'antica (claro, con los clarificaciones apropriadas $\partial \rightarrow \hat\partial$),
nosotros obtenemos otra vez que, en general, las derivadas {\it no} conmutan
$[{\hat\partial\over \hat\partial x_\mu},
{\hat\partial\over \hat\partial x_\nu}]_- \neq 0$.}

Adem\'{a}s, por raz\'{o}n de que la derivada en energ\'{\i}a corresponde al
operador de tiempo y la derivada en $i$-componente de momento, a $\hat{x}_{i}
$, vamos a proponer el siguente {\it anzatz} en la representaci\'{o}n de
momento lineal:
\begin{equation}
\lbrack \hat{x}^{\mu },\hat{x}^{\nu }]_{-}=\omega ({\bf p},E({\bf p}
))\,F_{||}^{\mu \nu }{\frac{\partial }{\partial E}}\,,\label{anz}
\end{equation}
con cierta funci\'{o}n del peso $\omega $ que sea diferente por elecciones
diferentes del base de spin para un campo antisim\'{e}trico tensorial.
La ecuaci\'on (\ref{anz}) puede sustituir el {\it anzatz} de H. Snyder.

En la literatura moderna, la idea de la rota invariancia de Lorentz por este
m\'{e}todo es concurrente con la idea de {\it la longitud fundamental}, por
primera vez introducida por V. G. Kadyshevsky~\cite{kadysh} (v\'{e}ase
tambi\'{e}n Ap\'{e}ndice A) a la f\'{\i}sica moderna en base a los antiguos
art\'{\i}culos de M. Markov. Ambas ideas y teor\'{\i}as correspondientes est\'{a}n
discutidos mucho, v\'{e}ase, por ejemplo,~\cite{amelino} (v\'{e}ase,
tambi\'{e}n Ap\'endice B). Personalmente, para mi la pregunta principal
es:  qu\'{e} es la escala espacial cuando la teor\'{\i}a de relatividad
empieza a ser incorrecta.

\bigskip
\bigskip

\noindent {\bf CONCLUSI\'ON}

\bigskip
\noindent Nosotros encontramos que el conmutator de dos derivadas puede ser
{\it no} igual a zero. Como consecuencia, por ejemplo, la pregunta sigue, si
la derivada $\hat{\partial}^{2}f/\hat{\partial}p^{\nu }\hat{\partial}p^{\mu }$
es igual a la derivada $\hat{\partial}^{2}f/\hat{\partial}p^{\mu }\hat{
\partial}p^{\nu }$ en todos los casos?\footnote{
La misma pregunta tambi\'{e}n aparece cuando tengamos diferenciaci\'{o}n con
respeto de los coordenadas, que puede tener impacto a los calculos
verdaderos del problema de una carga acelerada en electrodin\'{a}mica
clasica.} La consideraci\'{o}n presentada nos permite dar unos fundamentos
a las
teor\'{\i}as sobre espacios noconmutativas y induce a ver hacia el desarollo
futuro del an\'{a}lisis cl\'{a}sico para proveer una base matem\'{a}tica
rigurosa de operaciones con funciones que tienen ambos tipos de dependencia,
expl\'{\i}cita e impl\'{\i}cita.

\bigskip
\bigskip

\noindent {\bf AGRADECIMIENTOS}\newline

\bigskip \noindent Estoy agradecido a los profesores Dr. A. Chubykalo, Dr. L. M. Gaggero
Sager, Dr.  S. Vlayev, Lic. R. Flores Alvarado y los participantes del Seminario de la FF-UAZ,
donde esta idea apareci\'{o}, por discusiones. Los art\'{\i}culos viejos (que
aqu\'{\i} han sido citados) y discusiones viejas con sus autores fueron muy
\'{u}tiles para mi entendimiento de nuevas direcciones en f\'{\i}sica moderna.

\bigskip
\bigskip

\noindent {\bf AP\'{E}NDICE A. LA LONGITUD FUNDAMENTAL (INVARIANTE).}\newline

\bigskip \noindent Pienso que este concepto es equivalente a la
consideraci\'{o}n de la teor\'{\i}a de campos en el espacio de (anti) de Sitter.
\begin{equation}
p_{0}^{2}-p_{1}^{2}-p_{2}^{2}-p_{3}^{2}-M^{2}p_{4}^{2}=-M^{2}=-{\frac{1}
{\ell^{2}}}\,,
\end{equation}
donde $c=\hbar =1$, $M$ es arbitrario (por el momento) y esta conectado
con
el radio de {\it
curvatura}. El nuevo potencial electromagn\'{e}tico parece ser $5$-vector
que est\'{a} asociado con el grupo de Sitter $SO(4,1)$. Nuevas
transformaciones de norma intr\'{\i}nsecamente dependen de la longitud
fundamental ${\ell}$. Nuevas interacciones est\'{a}n mediadas por $\tau $
-photon (que es an\'{a}logo al concepto de Horwitz~\cite{Horwitz}). Existen
fermiones exoticos gracias al doblamiento del n\'{u}mero de componentes de
la funci\'{o}n de campo (espinores de ocho componentes):
\begin{mathletters}
\begin{eqnarray}
&&\left[ 2\sinh \mu/2-p_{\mu }\gamma ^{\mu }-(p_{4}-1)\gamma ^{5}\right] \Psi
(p,p_{4}) =0\, \\
&&\left[ 2\sinh \mu/2-p_{\mu }\gamma ^{\mu }+(p_{4}-1)\gamma ^{5}\right] \Psi
^{R}(p,p_{4}) =0.
\end{eqnarray}
\end{mathletters}
Estas ecuaciones tienen algo en com\'{u}n con las que fueron propuestas por
Tokuoka, SenGupta, Raspini y Dvoeglazov~\cite{gam}, $m=\sinh \mu $, $\sqrt{
1+m^{2}}=\cosh \mu =m_{4}$. En mis trabajos, las ecuaciones fueron derivadas en
base a los primeros principios (a la Sakurai). El procedimiento para obtener el
desplazamiento $e-\mu $ esta parecido al m\'{e}todo de Bogoliubov de rompimiento
de simetr\'{i}a en superconductividad.

\bigskip
\bigskip

\noindent {\bf AP\'{E}NDICE B. NO-CONMUTATIVIDAD.}\newline

\bigskip
\noindent Existen dos tipos de teor\'{\i}as noconmutativas:
\begin{itemize}

\item  El espacio-tiempo noconmutativo {\it can\'{o}nico} $[x_{\mu },x_{\nu
}]=i\theta _{\mu \nu }$;

\item  El espacio noconmutativo con \'{a}lgebra de Lie: $[x_{\mu },x_{\nu
}]=iC_{\mu \nu }^{\beta }x_{\beta }$.

\end{itemize}

Entre m\'{a}s estudiados est\'{a} el espacio $\kappa$-deformado de
Minkowski:
\begin{equation}
[ x_{m},t]={\frac{i}{\kappa }}x_{m}\,\,,\quad [ x_{m},x_{l}]=0\,.
\end{equation}
De otro modo fue demostrado que esta relacionado a las teor\'{\i}as
basadas en
\begin{equation}
E^{2}-c^{2}{\bf p}^{\,2}-c^{4}m^{2}+f(E,p,m,L_{p})=0\,,
\end{equation}
donde $L_{p}\sim\sqrt{{\frac{\hbar G}{c^{3}}}}\sim 1.6\cdot 10^{-35}\,m$.
Entonces, tenemos la escala de longitud que es independiente del observador
(sic!) J. Kowalski-Glikman y  G. Amelino-Camelia establecieron este mapeo y analizaron las conductas
asintoticas de momento lineal y energ\'{\i}a. En la escala de Planck, el
mundo puede ser {\it no}-relativista.

\end{document}